# Unconventional magnon transport in antiferromagnet NiPS$_3$ induced by an anisotropic spin-flop transition

Peisen Yuan[1*], Beatriz Martín-García[1,2], Evgeny Modin[1], M. Xochitl Aguilar-Pujol[1], Fèlix Casanova[1,2], and Luis E. Hueso[1,2*]

[1] CIC nanoGUNE BRTA, 20018 Donostia-San Sebastian, Basque Country, Spain
[2] IKERBASQUE, Basque Foundation for Science, 48009 Bilbao, Basque Country, Spain

**ABSTRACT**
Nonlocal magnon transport can provide valuable insight into the magnetic properties of magnetic insulators (MIs). A spin-flop transition, a typical magnetic reorientation in antiferromagnets, is expected to affect mag non transport, but studies on this topic are still rare and remain challenging, especially for van der Waals materials. Here we demonstrate the unconventional magnon transport driven by an anisotropic spin-flop transition in the van der Waals antiferromagnet NiPS$_3$. Examining the nonlocal voltage from thermally driven magnons reveals sharp jumps at certain directions when an in-plane magnetic field aligns with the *b*-axis of NiPS$_3$, attributed to an in-plane anisotropic spin-flop transition. Furthermore, thermally driven magnon signal exhibits a $1/d^2$ decay in thin NiPS$_3$, evidencing that it is dominated by the intrinsic spin Seebeck effect. Our findings highlight that the electrical detection of magnon currents in a nonlocal device geometry serves as a powerful approach for studying magnetic phase transitions in MIs.

The recent discovery of two-dimensional (2D) magnetic materials has attracted significant attention due to their very diverse magnetic properties and their potential for spintronic applications.[1-5] Among these materials, antiferromagnets show particular advantages for information storage due to their lack of stray magnetic fields and high-frequency dynamics.[6-9] Antiferromagnets also play a pivotal role in producing exchange bias, a key phenomenon in many modern spintronic devices.[7,10-15] Some antiferromagnetic materials can exhibit field-driven magnetic phase transitions. One of these is the so-called spin-flop transition, on which this work focuses. The magnetic field strength required for this transition ($H_{SF}$) to occur is closely related to both the effective exchange field strength ($H_E$) and the anisotropy field strength ($H_A$) within the antiferromagnetic material.[16-19] In 2D crystals, both interlayer and intralayer exchange interactions contribute to the overall magnetic behavior. This complex magnetic behavior in low-dimensional magnets still requires a deeper fundamental understanding before it can be exploited for spin-based applications.

Magnon transport experiments are a valuable tool for probing the pure spin response of 2D insulating materials. Magnon currents can be generated through spin pumping,[20,21] the spin Seebeck effect,[22-26] or electrical injection,[27-32] and are typically detected via the inverse spin Hall effect (ISHE) in nonmagnetic metals like Pt. Recent research studied the spin-flop transition in few-layer $MnPS_3$ using magnon transport.[33] This transition, however, can only be detected using Py via the anomalous SHE when an out-of-plane (OOP) magnetic field is applied, as the magnetic easy axis of $MnPS_3$ is OOP, perpendicular to the flake.[19,33-35] Similarly, $CrPS_4$, another 2D antiferromagnet with an OOP easy axis, shows a spin-flop transition at around 0.9 T under an OOP field.[36] However, magnon transport using Pt injectors and detectors cannot capture this low-field spin-flop transition.[31,37] For OOP 2D antiferromagnets, applying an OOP field is necessary to probe the spin-flop transition. The conventional ISHE only responds to in-plane (IP) spin polarization, making it unsuitable for detecting the spin-flop transition in these materials using Pt electrodes. $NiPS_3$ is an easy-plane antiferromagnet, where the spins predominantly align within the plane of the flake,[19,38,39] the standard geometry for detecting magnon propagation via the ISHE at a magnetic insulator (MI)/heavy metal interface should be appropriate.

In this work, we investigated the impact of the spin-flop transition on magnon transport in the easy-plane antiferromagnet $NiPS_3$. The experiments were performed by thermally exciting a magnon current in a few-layer-thick $NiPS_3$ flake, which was nonlocally detected by the ISHE using Pt contacts. The IP anisotropic spin-flop transition in $NiPS_3$ results in an unconventional angular dependence of the nonlocal signal. By exploring the field dependence of this nonlocal magnon signal, we were able to directly observe the anisotropy of the spin-flop transition in few-layer $NiPS_3$.

$NiPS_3$ belongs to the family of metal thiophosphates $MPX_3$ (M is a transition metal and X stands for a chalcogen) where Ni atoms are arranged in a honeycomb lattice structure in the *ab*-plane (Figure 1a). Bulk $NiPS_3$ exhibits XY-type antiferromagnetic ordering below its Néel temperature of around 150 K, with the major magnetic moments aligned in the *ab*-plane and a small portion of spins tilted out of plane.[38] Recent studies have identified IP multidomain structures in $NiPS_3$ flakes,[40,41] contributing to IP magnetic anisotropy.[42-46] Since the $H_{SF}$ is dependent on both the $H_A$ and $H_E$,[16-19,47] anisotropic $H_{SF}$ values are expected along different crystallographic axes of $NiPS_3$. Angle-dependent magnon transport under external magnetic field could reveal this anisotropic spin-flop transition, as demonstrated in our study. Although recent research has investigated IP anisotropy in $NiPS_3$ using neutron inelastic scattering and polarized light techniques, showing that the spins predominantly align along the *a*-axis,[43,46] these findings are insufficient to definitively classify $NiPS_3$ as a uniaxial antiferromagnet.[38,41,42,47] Therefore, in this work, the multidomain magnetic structure in $NiPS_3$, analogous to that in bulk antiferromagnet NiO films,[47,48] should be taken into account when considering the subsequent magnon transport. Bulk $NiPS_3$ possesses an energy gap of about 1.6 eV,[49] which leads to good insulating properties that enable the use of a nonlocal measurement configuration in our device. The crystallographic orientation of bulk $NiPS_3$ flakes on a Si/$SiO_x$ substrate (shown in Figure 1b) was initially investigated, with the *b*-axis determined using electron backscatter diffraction (EBSD) (Figure 1c). Subsequently, low-temperature polarized photoluminescence spectroscopy was conducted on the same flake (as detailed in Figure 1d and the experimental section). The photoluminescence spectrum, peaking at 1.475 eV (Figure 1e), reveals IP magnetic anisotropy due to its connection with exciton-antiferromagnetic coupling in NiPS$_3$, as seen in past works.[44,45] The polar plot of the polarized photoluminescence data (Figure 1f) shows a clear maximum at the angle corresponding to the *b*-axis of NiPS$_3$, as determined by EBSD. Consequently, a correlation between the crystallographic orientation and the IP anisotropy of NiPS$_3$ is established. In the subsequent magnon transport measurements using a nonlocal device structure, the *b*-axis of thin NiPS$_3$ flakes was verified using the same polarized-photoluminescence spectroscopy setup.

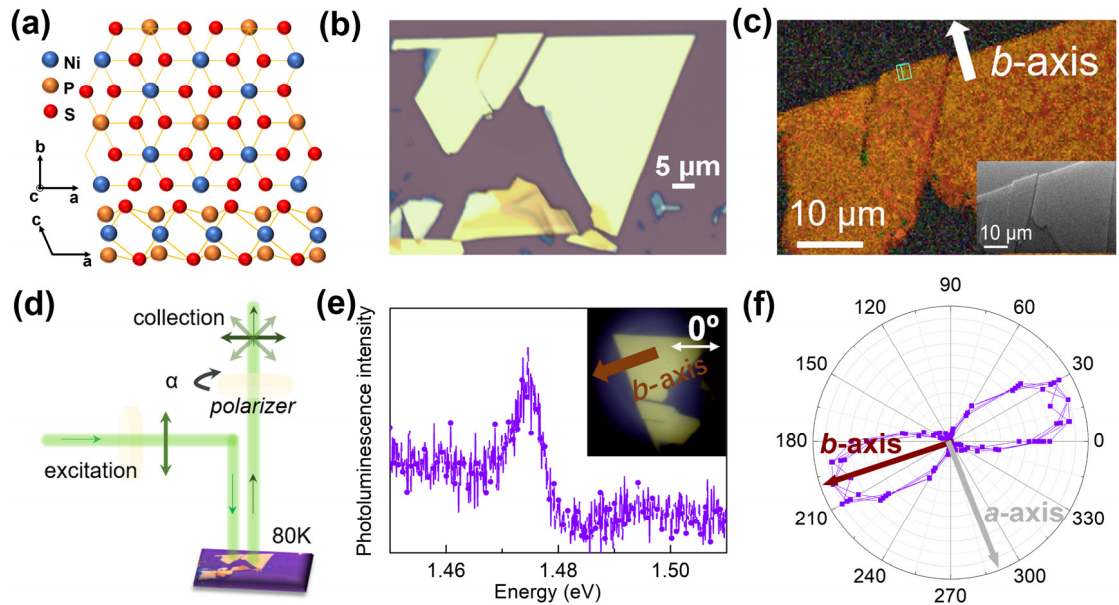

**Figure 1**. (a) Crystal and magnetic structure of NiPS$_3$. (b) Optical image of a bulk NiPS$_3$ flake used to determine the crystal orientation. (c) The inverse pole figure map measured by EBSD for the target NiPS$_3$ flake. The white arrow marks the *b*-axis of the NiPS$_3$ flake, and the unit cell is drawn in green. The inset shows the corresponding SEM image. (d) Schematic of the polarized photoluminescence measurement setup. The polarization in the excitation path is fixed, while the scattered light passes through a rotating polarizer and is collected as a function of the polarization angle α. (e) Representative photoluminescence spectrum collected at 80 K. The inset shows the relative orientation of the polarization and the *b*-axis (f) Photoluminescence intensity at the emission center as a function of the polarization angle α at 80 K. The *a* and *b* axes of the NiPS$_3$ flake are marked by gray and dark brown arrows, respectively.

For this work we fabricated devices from 12-nm- and 18-nm-thick NiPS$_3$ flakes. Combining optical and atomic force microscopy (AFM) data, we were able to calibrate the optical intensities of the different color channels for NiPS$_3$ flakes on Si/SiO$_x$ substrates viewed under an optical microscope, allowing us to accurately and nondestructively determine the thickness of the exfoliated NiPS$_3$ flakes (Figure S1a). The Raman spectra of the fabricated NiPS$_3$ devices show characteristic peaks consistent with those detected in bulk NiPS$_3$ (Figure S1b), indicating our fabrication process was successful. Figure 2a displays an optical image of one of the microfabricated nonlocal devices used to explore magnon transport in an 18-nm-thick NiPS$_3$ flake. Five differently oriented pairs of parallel Pt strips serve as injector-detector pairs for the magnon signals, with an injector-detector distance of 1 μm. The pair of strips approximately parallel to the *b*-axis is colored brown; the other four pairs with different alignments in the *ab* plane are colored pale yellow. The *b*-axis of this NiPS$_3$ flake was confirmed by polarized-photoluminescence spectroscopy (Figure S2a and S2b). For the nonlocal magnon devices, since the resistivity of the Pt is a critical factor for detecting magnon signals, the quality of the Pt deposition was assessed. This was done by measuring the resistivity of a Pt strip of another NiPS$_3$ device of the same thickness that was fabricated together with the one shown in Figure 2a. A four-point probe measurement yielded resistivity values in the expected range of 20-30 μΩ·cm (Figure S2c).

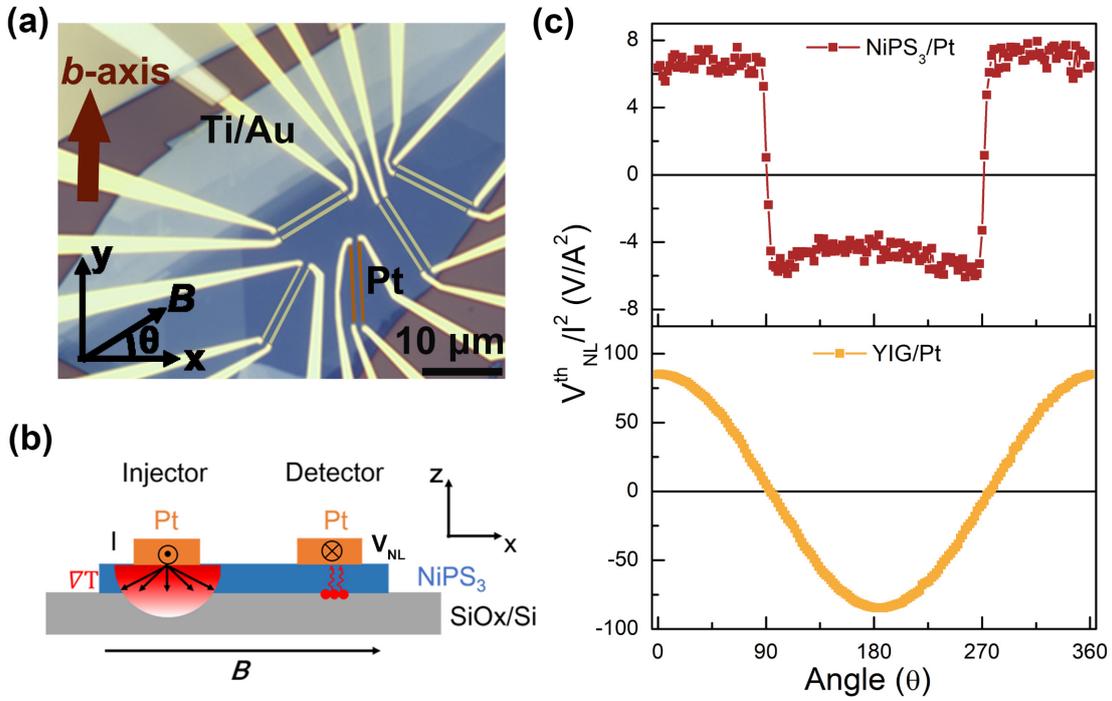

**Figure 2**. (a) Optical image of the fabricated nonlocal device. The parallel lines (brown and pale yellow) are the Pt strips, which are connected by Ti/Au electrodes (bright yellow); the blue underneath is the 18-nm-thick NiPS$_3$ flake. (b) Illustration of the device working process. An electrical current (*I*) is injected along a Pt strip, generating a radial thermal gradient (∇T). This thermal gradient creates a magnon population imbalance at both the Pt/NiPS$_3$ and NiPS$_3$/SiO$_x$ interfaces. A nonlocal magnon signal ($V_{NL}$) could be detected at another Pt strip, even within a far distance (*d*), due to the ∇T at the NiPS$_3$/SiO$_x$ interface, which induces magnon accumulation (small red balls) that diffuse toward the Pt detector. There, the magnon current enters the Pt detector as a spin current, which is converted into a transverse current and is detected as $V_{NL}$ in open circuit conditions. The external magnetic field *B* is an IP field, i.e., parallel to the NiPS$_3$ flake, and the angle between *B* and *I* is defined as θ, with θ = 0° when *B*

and *I* are perpendicular (c) Above: Angular dependence of the nonlocal thermally excited magnon signal in an 18-nm-thick NiPS$_3$ flake with two parallel Pt strips oriented close to the *b*-axis of NiPS$_3$ flake. Below: for comparison, signal recorded on a sample with an identical geometry but patterned on YIG. The measurements were performed at 20 K with *B*= 9 T and *I*=100 µA. The thermally driven magnon signals are normalized to the square of the injected current ($V_{NL}^{th}/I^2$). Angle-dependent second-harmonic local voltage measurement of a Pt strip is shown in Figure S3.

Generally, electrically and thermally excited magnon spin currents can be generated in Pt/MI systems by passing a current through the Pt strips.[24] This current induces spin accumulation at the Pt/MI interface via the SHE,[50-52] creating an imbalance of the magnon population in MI and leading to magnon diffusion within MI, with a characteristic relaxation length ($\lambda_m$). The electrically excited magnons can then be detected at another Pt strip through the ISHE as a nonlocal first-harmonic voltage. On the other hand, the current flowing through the Pt injector also generates a radial temperature gradient in MI due to Joule heating. This radial thermal gradient generates a negative magnon chemical potential ($\mu_m^-$) near the Pt heater and a positive one ($\mu_m^+$) farther from the Pt heater, driving a magnon current in MI. This magnon current can be detected at a second Pt strip via the ISHE as a nonlocal second-harmonic voltage. Due to the radial distribution of temperature, the $\mu_m^+$ at positions further from the Pt heater initially increases, followed by lateral diffusion along the MI layer, and decaying exponentially over a characteristic length scale, denoted as $\lambda_m$. Consequently, $\lambda_m$ can be extracted by fitting the nonlocal harmonic signals within the exponential decay regime. Notably, as extensively studied in YIG systems,[53,54] when the Pt-Pt distance d≥ 3-5 $\lambda_m$, where $\mu_m^+$ from the Pt heater is nearly zero, a nonlocal magnon signal remains detectable due to the ∇T at the MI/substrate interface. In this regime, magnons exhibit a $1/d^2$ decay, characteristic of the intrinsic spin Seebeck effect, which is further demonstrated to be present in the NiPS$_3$ system (shown in Figure 2b). For NiPS$_3$, a sizeable spin signal was detected at the Pt/NiPS$_3$ interface through spin Hall magnetoresistance in a recent study, demonstrating effective spin-current conversion at the surface.[39] In this work, we focus on the nonlocal second-harmonic voltages only, i.e., the thermally excited magnon signal ($V_{NL}^{th}$), since no first-harmonic voltage could be detected. The possible reason for not observing the first harmonic signal is discussed in the Supporting Information.

We first measured the angle-dependent thermally excited magnon signals in Pt/NiPS$_3$ under a 9 T IP magnetic field at 20 K (Figure 2). Our results, shown in Figure 2c, are quite remarkable: the thermally excited magnon signal shows an unconventional angular dependence when the Pt strips are oriented close to the *b*-axis of the NiPS$_3$ flake, presenting sharp jumps at certain values of θ, in stark contrast to the typical *sin* function observed in well-known systems such as yttrium iron garnet (YIG)/Pt,[24] also shown for comparison. This behavior was still observed at other temperatures and injected currents (Figure S4). Additionally, the magnon amplitude in the antiferromagnet NiPS$_3$ is significantly smaller than that in YIG systems. Previous reports suggest that magnon transport in antiferromagnetic insulators (AFIs) may involve two degenerate magnon modes with opposite chirality, which typically cancel each other, leading to zero spin current transport.[55-57] To generate finite spin signals, a sufficiently strong magnetic field is required to break the degeneracy of these magnon modes. However, even then, the effect is weak, producing only small magnon signals.

Regarding the unconventional angle-dependent second harmonic signal, a plausible explanation is the sudden change in the spin configuration induced by the spin-flop transition in $NiPS_3$. It is worth noting that the spin-flop transition has been previously observed in bulk $NiPS_3$ by conventional magnetometry but has not been reported for few-layer flakes until now.[19] One reason is that the magnetic structure of thin-layer $NiPS_3$ is unstable and exhibits multidomain states.[40,41] Since the spin-flop transition depends on the strength of the applied magnetic field, we expect the angular dependence of thermally excited magnons to vary with decreasing field strength, as confirmed in Figure 3a. The curves display a clear jump behavior above 7 T but tend toward a *sin*-like shape at lower fields. In previous reports on easy-axis antiferromagnets with single-domain structures,[58] nonlocal thermal magnon signals are proportional to the component of the net magnetization vector (*m*) normal to the Pt strips. Below the spin-flop transition, the Néel vector (*n*) remains aligned along the easy-axis direction, resulting in zero net magnetization normal to the Pt strip, and thus, no nonlocal thermal magnon signal is generated. In contrast, the influence of multidomain structures in this easy-plane antiferromagnet $NiPS_3$ flake under varying magnetic fields must be considered, as observed in similar easy-plane bulk NiO films.[47,48] This will be further discussed below.

The photoluminescence peak at around 1.475 eV in Figure 1e is demonstrated to reflect the information of antiferromagnetic ordering of the $NiPS_3$ so the detected clear angle-dependent photoluminescence spectra in Figure 1f and Figure S2b indicate that these $NiPS_3$ flakes exhibit a certain IP magnetic anisotropy.[44,45] The $H_{SF}$ values of antiferromagnets are proportional to the strength of $H_E$ and $H_A$, where $H_A$ is related to both the intrinsic magnetic anisotropy and the additional anisotropy contributed by the external field.[13] Therefore, the IP magnetic anisotropy and the different initial angles between the external magnetic field and the crystal orientation of the $NiPS_3$ flake are expected to induce anisotropic $H_{SF}$, which is the direct cause of the unconventional nonlocal magnon signals.

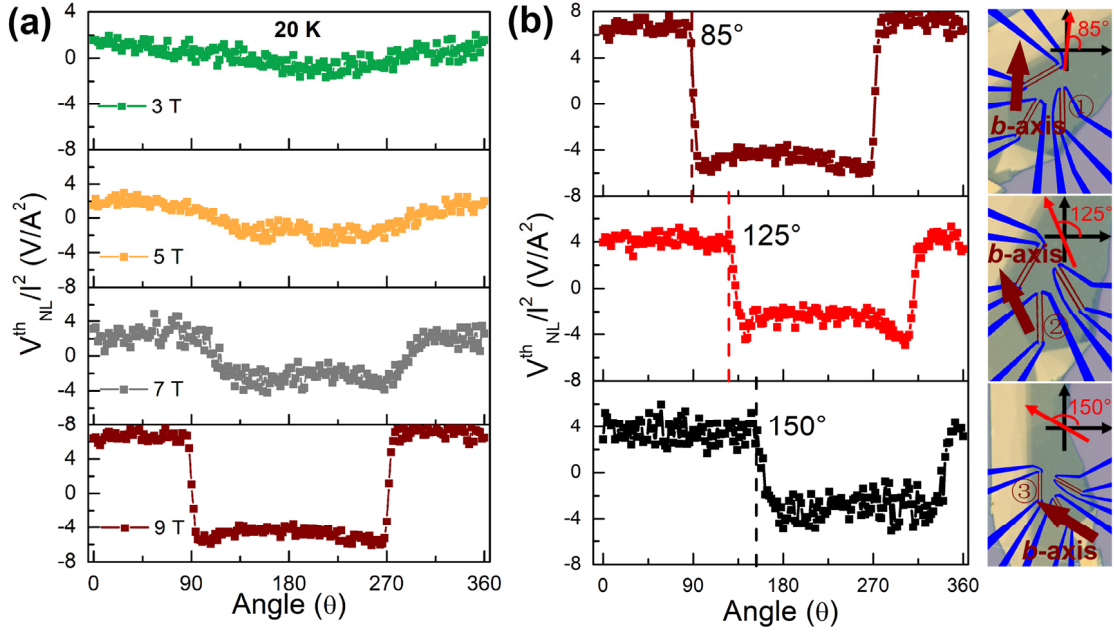

**Figure 3**. (a) Angular dependence of the nonlocal thermally excited magnon signals measured at different external magnetic fields. (b) Comparison of the angular dependence of the nonlocal thermally excited magnon signals with $B$= 9 T in three devices oriented at different angles with respect to the *b*-axis of the NiPS$_3$ flake. The angle at which the spin-flop transition is observed is labeled. The corresponding device images are shown on the right. All measurements were performed at 20 K with an applied electric current through the Pt injector of 100 μA. The Pt injector-detector distance is 1 μm.

To further understand the effect of this anisotropic H$_{SF}$ on the unconventional magnon transport, we fixed the external magnetic field strength at 9 T and measured the angle-dependent second harmonic signals in two other nonlocal Pt structures (Figure 3b), with angles of around 30° (device 2) and 60° (device 3) with respect to the *b*-axis of NiPS$_3$. The unconventional nonlocal magnon transport behavior is still observed in these devices (Figure 3b), but the jump related to the spin-flop transition shifts from 85° (device 1) to 125° (device 2) and 150° (device 3), where we marked these directions as red arrows. Importantly, all these arrows are similarly aligned to the *b*-axis of the NiPS$_3$ flake. These results indicate that, near the *b*-direction of NiPS$_3$ at an external magnetic field strength of 9 T, the spin-flop transition is strongly activated (H$_{SF}$<9 T), but does not occur when the field direction deviates from the *b*-axis, in which case a higher field strength is required to produce the spin-flop transition (H$_{SF}$>9 T). Based on the reported multidomain magnetic structures,[40,41] the IP magnetic anisotropy,[44,45] and the jump behavior of nonlocal magnon signals in this work, we propose the existence of two magnetic domains with spin configurations aligned along the *a*-axis and *b*-axis of NiPS$_3$, respectively, to understand our experimental results.

A schematic diagram is proposed in **Figure 4a** to illustrate the experimentally observed unconventional angle-dependent nonlocal magnon signal. We focused on the high-field (9 T) case, considering the contributions from two domains in NiPS$_3$: one with spins aligned along the *a*-axis (domain 1), which has been previously demonstrated,[44,45] and the other with spins along the *b*-axis (domain 2), differing from recent studies on multidomain structures in pristine NiPS$_3$ flakes.[40,41] The origin of the proposed *b*-axis magnetic domain structure is discussed in the Supporting Information.

When *B*//*a*-axis, the 9 T field is insufficient to induce a spin-flop transition in domain 1. Instead, it causes a slight coherent rotation of the *n* vector (middle panel of Figure 4a),[59] deviating from the field direction and contributing minimally to the nonlocal thermal magnon signals. Simultaneously, since *B* is perpendicular to the spins in domain 2, the *m* vector is generated normal to the Pt strips due to the canting effect, contributing to the nonlocal thermal magnon signals. As reported previously, the *m* vector increases linearly with increasing magnetic field, resulting in a corresponding linear increase in the nonlocal thermal magnon signals,[58] consistent with our field-dependent measurements (Figure 4c).

For *B*//*b*-axis, the *m* vector in domain 1 is nearly parallel to the Pt strips, leading to approximately zero nonlocal thermal magnons. Based on the observed near-zero nonlocal magnon signals close to 90° (Figure 4b), we infer that a spin-flop transition occurs in domain 2, since only in this case would the spin-flop-induced *m* vector align along the Pt strips, resulting in zero nonlocal thermal magnon signals. Indeed, a sharp change with a small amplitude is observed in the field-dependent nonlocal magnon curve when *B* is aligned along the *b*-axis, corresponding to the spin-flop transition (Figure 4d). Notably, the small detected nonlocal signal is attributed to a slight angular offset (2–3°) between the Pt strips and the *b*-axis during device fabrication, which introduces a component of the *m* vector perpendicular to the Pt strips. At other angles, the contribution of the total *m* vectors induced by both domains should be considered in determining the final nonlocal thermal magnon signals. The same conclusions are further supported by the investigation of angle-dependent and field-dependent nonlocal magnon currents in a device (Device 4) with Pt strips aligned along the *a*-axis of NiPS$_3$ (Figure S5). The lower-field case is discussed in the Supporting Information.

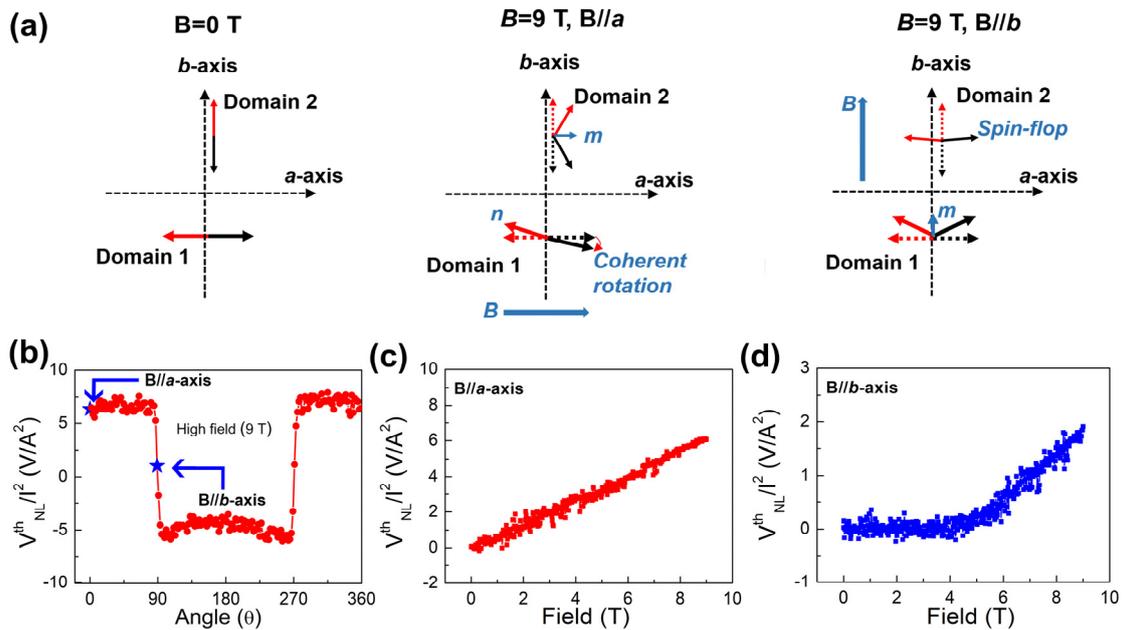

**Figure 4**. (a) Different spin configurations occurring rotating the angle between the external magnetic field and the NiPS$_3$ flake at high field (9 T). Two domain structures, one aligned along the *a*-axis (domain 1) and the other along the *b*-axis (domain 2), are considered. A high magnetic field applied along the *a*-axis induces a coherent rotation of the Néel vector (*n*) in domain 1, while a field applied along the *b*-axis triggers a spin-flop transition in domain 2. In the middle and right panels, the dashed arrows represent the initial spin orientation of the two domains, while the solid arrows indicate their final spin states at 9 T. (b) Angular dependence of the nonlocal thermally excited magnon signals measured at 9 T, where the blue star points stand for the magnetic field directions along the *a*-axis and *b*-axis of NiPS$_3$, respectively. (c)-(d) Field-dependent nonlocal magnon signals for the magnetic field applied along the *a*-axis (c) and *b*-axis (d) of the NiPS$_3$ flake, where an upturn corresponding to the spin-flop transition is clearly observed when B//*b*-axis.

To further investigate the decay of thermal magnons in relation to the spin-flop transition in NiPS$_3$, we measured the angle-dependent thermally excited magnon signals for Pt-Pt separations ranging from 0.8 to 3.8 µm in a thinner 12-nm NiPS$_3$ flake (**Figure 5**). The unconventional angular dependence is still observed for all injector-detector distances, indicating that our results are general and, as expected, they do not change with the thickness of NiPS$_3$. We plotted the amplitudes of the magnon currents as function of *d* (Figure 5c) and found it does not follow the exponential decay typically observed in many ferromagnetic or AFI. In Figure 5c, the data is well fitted by the $1/d^2$ function (the green solid line) but not by the general exponential function for the magnon diffusion (the red solid line).[27] As discussed, the $1/d^2$ decay regime of the nonlocal signal induced by thermally driven magnons has been studied well in YIG,[53,54] dominating at distances $d \geq 3$-5 $\lambda_m$. In YIG, with $\lambda_m$ on the order of tens of µm, *d* must be quite large (>50 µm) to observe the $1/d^2$ decay region. However, the situation is quite different in 2D antiferromagnets, which typically exhibit much shorter $\lambda_m$ (e.g., $\lambda_m \approx 1$ µm for 8-nm-thick MnPS$_3$, $\lambda_m \approx 0.5$ µm for 20-nm-thick FePS$_3$, $\lambda_m \approx 0.2$-0.8 µm for 160-nm-thick CrPS$_4$) compared to YIG,[24,60,61] with further reductions at higher temperatures and in thinner samples. In our case, $\lambda_m$ value has also been detected in different thickness of NiPS$_3$ flake (Figure S6). It shows a maximum value of around 1.1 µm in thicker devices but will decrease fast with reducing the thickness. In the 12 nm-thick NiPS$_3$, the $\lambda_m$ value cannot be extracted by the typical exponential fitting. It means that the $\lambda_m$ in such thin NiPS$_3$ should be very short, which enables us to observe the intrinsic spin Seebeck effect in the Pt-Pt distance range from 0.8 µm to 3.8 µm and exhibits the $1/d^2$ decay behavior of the thermally driven magnons (as stretched in Figure 2b).

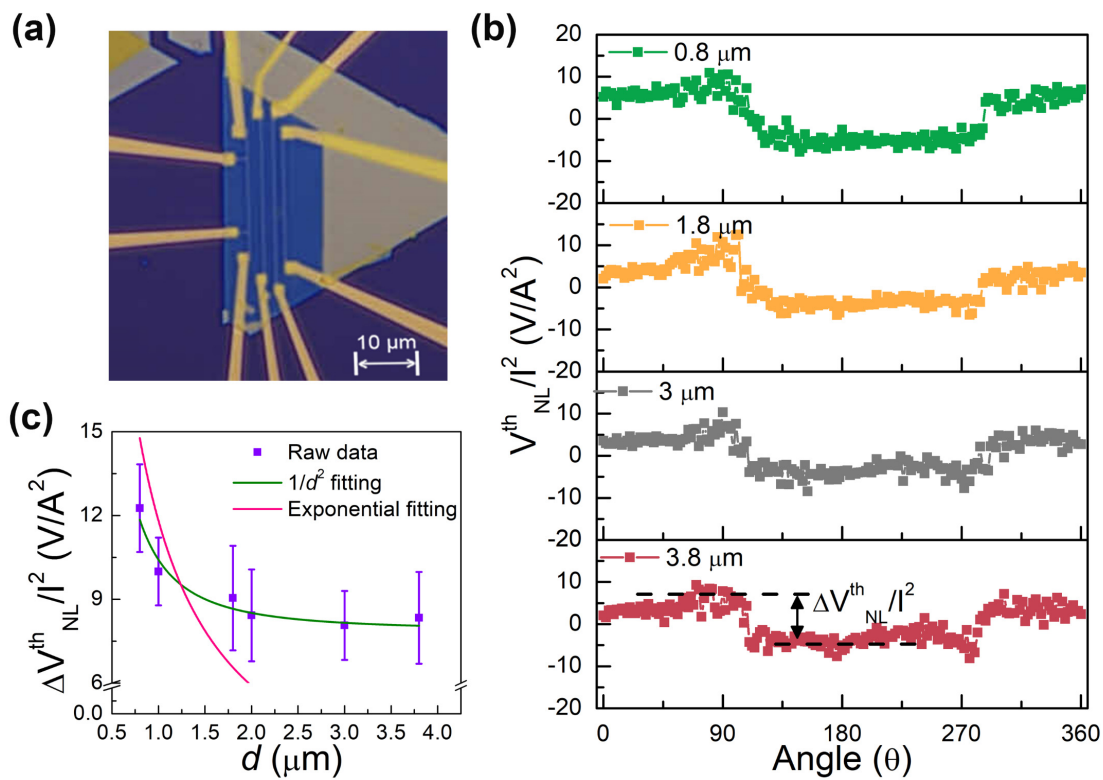

**Figure 5.** (a) Optical image of the fabricated nonlocal device with a 12-nm-thick NiPS$_3$ flake. (b) Angular dependence of the nonlocal thermally excited magnon signal of the device shown in panel (a) for different injector-detector distances $d$. The measurements were performed at 20 K, 9 T and 100 μA. (c) Amplitude of the angle-dependent nonlocal magnon signal as a function of $d$, where the amplitude is extracted from the difference between the maximum and minimum values (two dashed lines in the bottom panel of Figure 5b). Solid lines are different fits to the data.

In summary, we investigated thermally excited magnon transport in the 2D AFI NiPS$_3$ using Pt contacts in a nonlocal device geometry. The angle-dependent nonlocal magnon signal observed in this material exhibits sharp jumps caused by the sudden reorientation of the spins during the spin-flop transition. The results indicate that, at high magnetic field, two domain structures with spins aligned along the *a*-axis and *b*-axis of NiPS$_3$ contribute to the nonlocal thermally driven magnon signals. Additionally, we observed that the nonlocal signals exhibit a $1/d^2$ decay in a thin NiPS$_3$ device, possibly due to the contribution of the intrinsic spin Seebeck effect. Finally, our field-dependent measurements of the nonlocal magnon signal serve as direct verification of the in-plane anisotropy of the spin-flop transition in NiPS$_3$.

**Supporting Information**
The Supporting Information is available free of charge at http://pubs.acs.org.
Detailed description of the sample fabrication and characterization, discussion about the first harmonic nonlocal signal, the origin of the proposed domain structure with spin along the *b*-axis of NiPS$_3$, and the nonlocal magnon signals at low-field case, additional experimental results, including the characterization of NiPS$_3$ flakes by AFM and Raman spectroscopy, angle-dependent PL spectrum of the 18 nm-thick NiPS$_3$ magnon device, the resistivity of Pt electrodes, angle-dependent local voltage measurement from a Pt strip, angle-dependent nonlocal magnon signals at other temperatures and injected currents, angle-dependent and field-dependent nonlocal magnon currents with Pt strips along the *a*-axis of NiPS$_3$, magnon diffusion length in different thickness of NiPS$_3$ devices, Raman spectra measured from pristine NiPS$_3$ and NiPS$_3$ with deposited Pt.

**Acknowledgements**

We acknowledge financial support from the Spanish MICIU/AEI/10.13039/501100011033 (CEX2020-001038-M), MICIU/AEI and ERDF/EU (PID2021-122511OB-I00) and by the European Union's Horizon 2020 Research and Innovation Programme (Grant Agreement Nos. 965046-INTERFAST and 964396-SINFONIA). P.Y. and B.M.-G. acknowledge support from MICIU/AEI and European Union NextGenerationEU/PRTR (grant Nos. FJ2020-044666-I and RYC2021-034836-I, respectively). M.X.A.-P. thanks the Spanish MCIU/AEI for a Ph.D. fellowship (grant No. PRE-2019-089833).

**TOC Graphic**

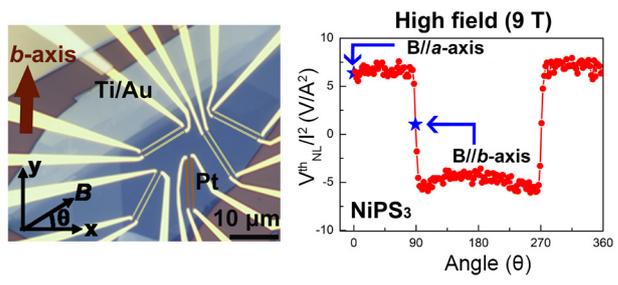

# Supporting Information

# Unconventional magnon transport in antiferromagnet NiPS$_3$ induced by an anisotropic spin-flop transition

Peisen Yuan[1*], Beatriz Martín-García[1,2], Evgeny Modin[1], M. Xochitl Aguilar-Pujol[1], Fèlix Casanova[1,2], and Luis E. Hueso[1,2*]

[1] CIC nanoGUNE BRTA, 20018 Donostia-San Sebastian, Basque Country, Spain
[2] IKERBASQUE, Basque Foundation for Science, 48009 Bilbao, Basque Country, Spain

# Experimental Section

*Sample preparation.* Few-layer flakes of NiPS$_3$ were exfoliated in Ar-filled glovebox from a bulk NiPS$_3$ crystal (HQ Graphene) and transferred onto a Si/SiO$_x$ (300 nm) substrate. We selected two different thicknesses of NiPS$_3$, 12 nm and 18 nm, to fabricate two series of samples. Considering the stability of monolayer NiPS$_3$ in air condition for few days, (1) the nonlocal devices were able to fabricated outside the glovebox within just few hours by e-beam lithography, using a positive resist of poly(methyl methacrylate) (PMMA) to pattern the mask. For the Pt contacts, 200 nm × 13 μm strips were first patterned on the PMMA. E-beam lithography was then performed, followed by the deposition of 5-nm of Pt by high-vacuum magnetron sputtering (3×10$^{-8}$ torr), and finally the lift-off process. After that, Ti (5nm)/Au (45nm) contacts were fabricated following the same procedure. After the process, the samples are always kept inside the Ar-filled glovebox.

*Scanning electron microscopy and electron diffraction characterization of flakes.* The crystallographic orientation of NiPS$_3$ flakes was measured by employing electron backscatter diffraction (EBSD). The measurements were taken in a Tescan AmberX SEM (Tescan, Czech Republic) equipped with an EDAX Hikari EBSD camera (EDAX, USA). The microscope was operated at an accelerating voltage of 20 kV and electron beam current of 10 nA. The orientation map was collected from an area of 58 μm × 39 μm with a step size of 0.3 μm.

*Low-temperature polarized photoluminescence spectroscopy characterization.* The NiPS$_3$ flakes' photoluminescence intensity measurements (peak center ca. 1.47 eV at 80 K) varying the polarization of the incident laser and/or collection were carried out on a Renishaw® inVia Qontor micro-Raman instrument equipped with a Linkam® vacuum chamber coupled to a liquid N$_2$ cooling system. We collected the photoluminescence spectra using a 532 nm laser as excitation source, a diffraction grating of 1800 lines/mm and a 50× objective (W.D. 11 mm). The incident laser power was kept < 3 mW to avoid damage to the flakes during acquisition. Angle-dependent linearly polarized photoluminescence measurements were carried out using two configurations. In the first one, configuration I, we used a motor (Thorlabs, ELL14K Rotation Mount – PC controlled by ELLO® software) to rotate a half-wave plate (Thorlabs, AHWP10M-580) leading to the change of polarization in the excitation path while keeping a linear polarizer (Renishaw®) fixed in the detection path. At the starting point, 0 degrees, both polarizations in excitation and collection are parallel. In this way, we observe that the photoluminescence emission intensity is isotropic with respect to the rotation angle of the polarization in the excitation path (circular polar plot). In the second configuration, configuration II, we keep the polarization in the excitation path fixed while rotating a linear polarizer (Thorlabs, LPVIS100-MP2) with a motor (Thorlabs, ELL14K Rotation Mount – PC controlled by ELLO® software) to vary the polarization in the detection path. At the starting point, 0 degrees, both polarizations in excitation and collection are parallel. In this way, we observe that the photoluminescence emission intensity is maximum with respect to the rotation angle of the polarization in the collection path (polar plot with two maxima) along the *b*-axis of the flake, determined by EBSD-SEM measurements. As control experiments, the angle-dependent polarized-response of the Si peak intensity at 520.5 cm$^{-1}$ in Si/SiO$_2$ substrates (100) was checked.

*Optical and electrical characterization.* The relationship between the thickness of the NiPS$_3$ flakes and the optical intensity was calibrated by atomic force microscopy (Agilent 5500 SPM) and optical microscopy. Raman measurements of the fabricated device were carried out using the previously mentioned Renishaw® inVia Qontor micro-Raman instrument with a 100× objective and an excitation wavelength of 532 nm. For the electrical measurements, the sample was loaded into a physical property measurement system (PPMS) by Quantum Design. A Keithly 6221 was used to apply an AC current through the Pt injector and a nonlocal voltage signal ($V_{NL}$) was measured at the Pt detector by a Keithley 2182 nanovoltmeter. In order to obtain the thermally excited magnon signal, we applied the injected current (i) under a delta mode (i+ 0, i- 0) to extract the second harmonic term from the equation: $V_{NL}$=($V_{NL}$(i+ 0)+ $V_{NL}$(i+ 0))/2. The applied magnetic field was applied in the plane of the device.

## Discussion

### The absence of the first harmonic signal

According to several studies on magnon transport in 2D antiferromangets, such as $MnPS_3$, $FePS_3$, and $CrPS_4$, (2-4) all of them reported the absence of the first harmonic signal, even employing highly sensitive measurement techniques. One proposed explanation from these studies is that the first harmonic signal ($V_{NL}^{1\omega}$) decreases significantly as the temperature is reduced. For example, in YIG it is difficult to detect below 10 K. (5) However, in our study, even with temperatures raised to 40 K, the first harmonic signal remains absent. These findings suggest that the absence of $V_{NL}^{1\omega}$ is not solely due to measurement setups but is strongly related to the studied system. Particularly in 2D systems, there may be distinct mechanisms affecting nonlocal magnon transport compared to 3D systems. Further investigation and dedicated efforts are necessary to deeply explore and understand these phenomena in the future.

### The origin of the proposed domain structure with spin along the *b*-axis of NiPS$_3$

The proposed another domain structure with spins along the *b*-axis differs from recent studies reporting multidomain structures in pristine NiPS$_3$ flakes by Tan et al. and Lee et al., (6,7) as our work explores the magnetic texture within a device structure. However, our assumption is reliable and expected based on the results from Tan et al. and Lee et al.

According to the studies by Tan et al. and Lee et al., a key finding is that the detected magnetic texture in NiPS$_3$ is strongly dependent on both its thickness and thermal fluctuations. In Tan's work,(6) magnetic domains were mainly investigated in thin NiPS$_3$ flakes (fewer than four layers) at low temperatures. In this case, due to the strong local strain effect, three distinct domain structures—with spins oriented along the *a*-axis and ±120° offset from the *a*-axis— were detected in pristine NiPS$_3$ using the linear dichroism (LD) technique. The domain distribution varies with temperature and thickness due to competition between strain effects and thermal fluctuations, leading to magnetic domain flipping and the emergence of different domain structures.

Due to the limited spatial resolution of optical LD measurements (on the micrometer scale), smaller domain structures could not be effectively observed in Tan's study. However, Lee et al., using X-ray magnetic linear dichroism (XMLD) with higher spatial resolution, identified additional smaller domains (on the order of hundreds of nanometers) with spin orientations offset by 10–30° from the *a*-axis of NiPS$_3$. (7) Since Lee et al. focused on thicker NiPS$_3$ flakes, their detected dominant domain structure was aligned along the *a*-axis, likely because the local strain effect was negligible. Consequently, they did not observe the ±120° rotated domains reported by Tan et al.  However, their study also demonstrated that these small magnetic domains are strongly influenced by thermal fluctuations.

Based on the studies by Tan et al. and Lee et al., the proposed magnetic domain orientation along the *b*-axis of NiPS$_3$ under the Pt electrodes in our work can be understood. As demonstrated in our previous study, (8) Pt deposition induces a disordered contact region approximately 2–3 nm thick on the surface of the layered two-dimensional material, leading to significant distortion of the contact region of the NiPS$_3$ flakes and causing a large strain effect. This strain-induced alteration of NiPS$_3$ can be confirmed by examining the Raman spectrum, a useful tool for probing surface atomic ordering (as shown in the Figure S7). It is noteworthy that, after Pt deposition, while the main characteristic peaks remain detectable, certain peaks—such as the Eg$^{(3)}$ and Eg$^{(4)}$ modes—disappear completely, indicating that the surface atomic ordering in NiPS$_3$ is indeed modified by the Pt deposition.

Building on Tan's work, the strain effect induced by the Pt deposition can significantly influence the distribution of magnetic domains. This effect may lead to the emergence of other domains with spin rotations of approximately 120° (or -120°) along the *a*-axis of NiPS$_3$. Additionally, we must account for the impact of thermal fluctuations on domain distribution, especially since injected currents in the Pt wire generate Joule heating. As demonstrated by Lee et al., thermal fluctuations can further modify the rotation of the spins, causing small-size magnetic domains to shift by 10–30°. Considering the narrow Pt strip (300 nm in width) in our device structure, it is reasonable to expect such a small magnetic domain under the Pt strip that may have spins aligned approximately 90° (along the *b*-axis), offset from the *a*-axis of NiPS$_3$. This expectation aligns with both the strain and thermal fluctuation effects observed in the works of Tan et al. and Lee et al. However, detecting such a small-scale magnetic domain structure at the Pt/NiPS$_3$ interface is challenging using traditional polarized photoluminescence (PL) or LD techniques, as the metallic Pt layer reflects a portion of the polarized light, introducing measurement artifacts. Instead, the Néel vector orientations within these domains can be inferred from our spin-related magnon transport measurements, highlighting the importance of this approach in probing spin textures that are otherwise difficult to access.

**The nonlocal magnon currents at low-field case**

At lower magnetic fields, the magnetic structure of NiPS$_3$ becomes more complex due to the presence of low-energy domains. When the Zeeman energy induced by the magnetic field is sufficient to lift the degeneracy of these energetically equivalent domains, even a small external field can alter the spatial distribution of antiferromagnetic domains by pushing domain walls into energetically unfavorable regions. This redistribution of domains-similar to that in ferromagnetic or ferrimagnetic systems-with various angles between the applied magnetic field and the spin orientations of the domains, primarily contributes to the nonlocal magnon signal, resulting in a *sin*-like angle-dependent function. However, the overall magnitude of the signal remains weak due to compensation effects, although we still observe some weak magnon signals for fields below the spin-flop field.

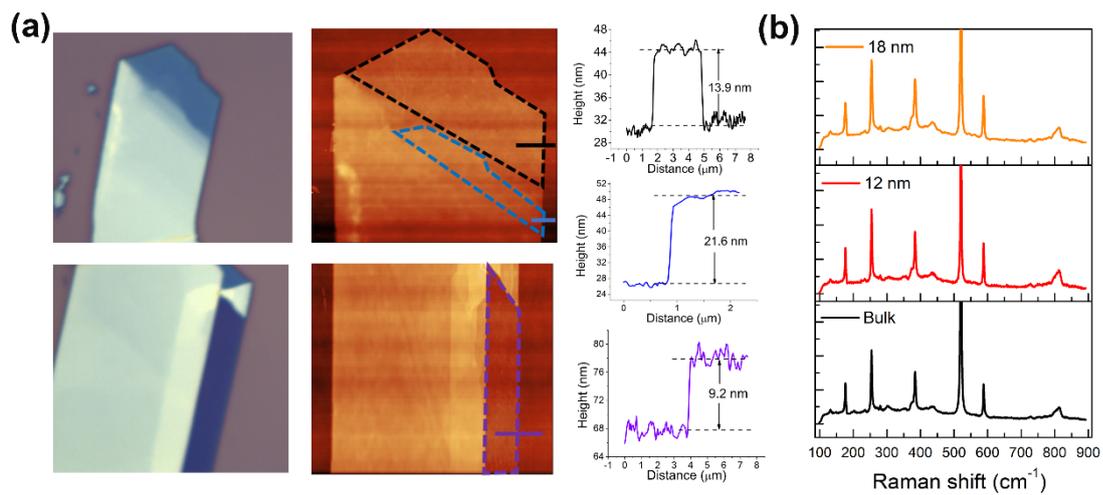

**Figure S1**. (a) Optical image of exfoliated NiPS$_3$ with different optical color and the measured thickness by AFM. (b) Representative Raman spectra of the NiPS$_3$ flakes in the fabricated devices.

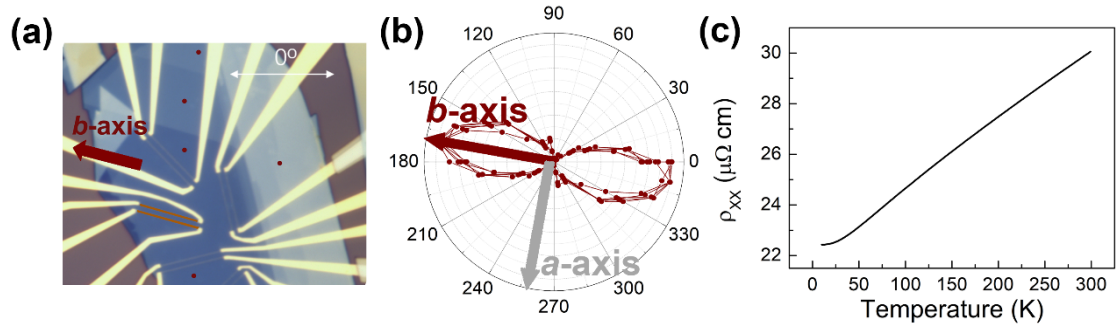

**Figure S2**. (a) Optical image and (b) the corresponding angle-dependence of the photoluminescence intensity of the 18 nm-thick NiPS$_3$ device using configuration II collected in 5 different points at 80 K. The *a*-axis and *b*-axis of this NiPS$_3$ flake are marked by gray and dark brown arrows, respectively. (c) Longitudinal resistivity of the Pt strips as function of temperature.

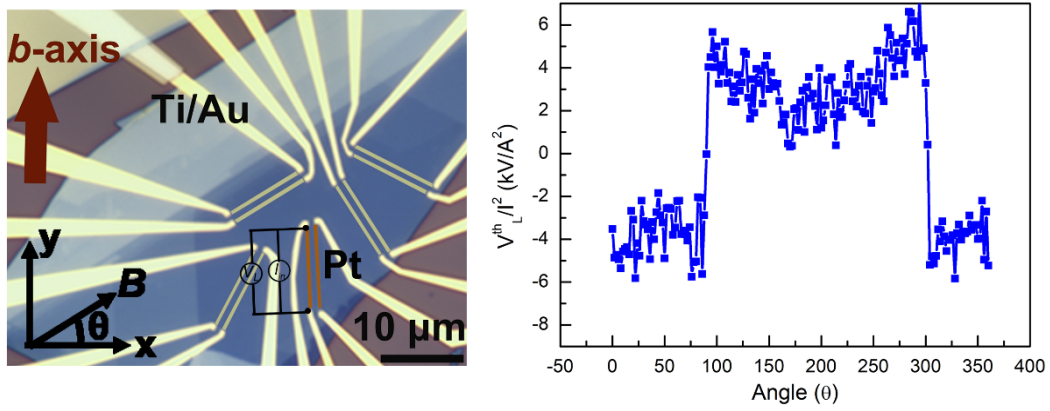

**Figure S3**. Angle-dependent second-harmonic local voltage ($V_L^{th}$) measurement of the Pt strip (as shown in the left image). The measurements were performed at 20 K with $B$= 9 T and $I$=100 µA. The signal is normalized to the square of the injected current ($V_L^{th}/I^2$).

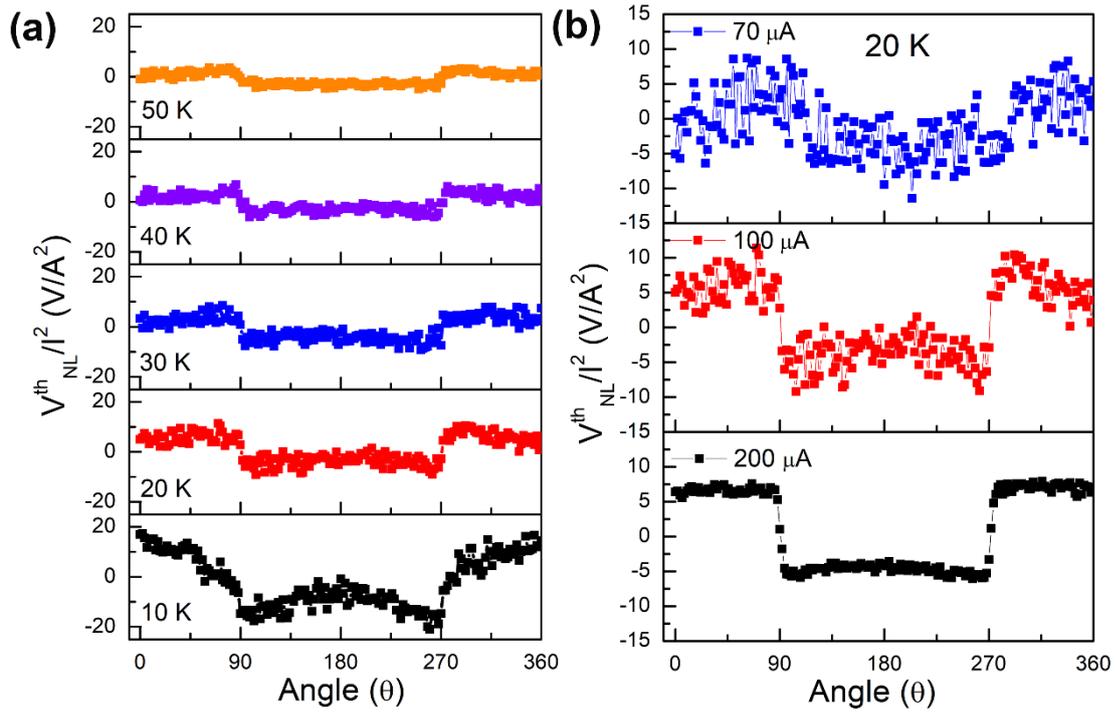

**Figure S4**. (a) Angle-dependent nonlocal magnon signal under thermal excitation at different temperatures. The device was measured with an injected electric current of 100 μA and an applied magnetic field of 9 T. (b) Angle-dependent nonlocal magnon signal under thermal excitation for different values of the electric current injected through the Pt injector. The applied magnetic field is 9 T and the temperature is 20 K.

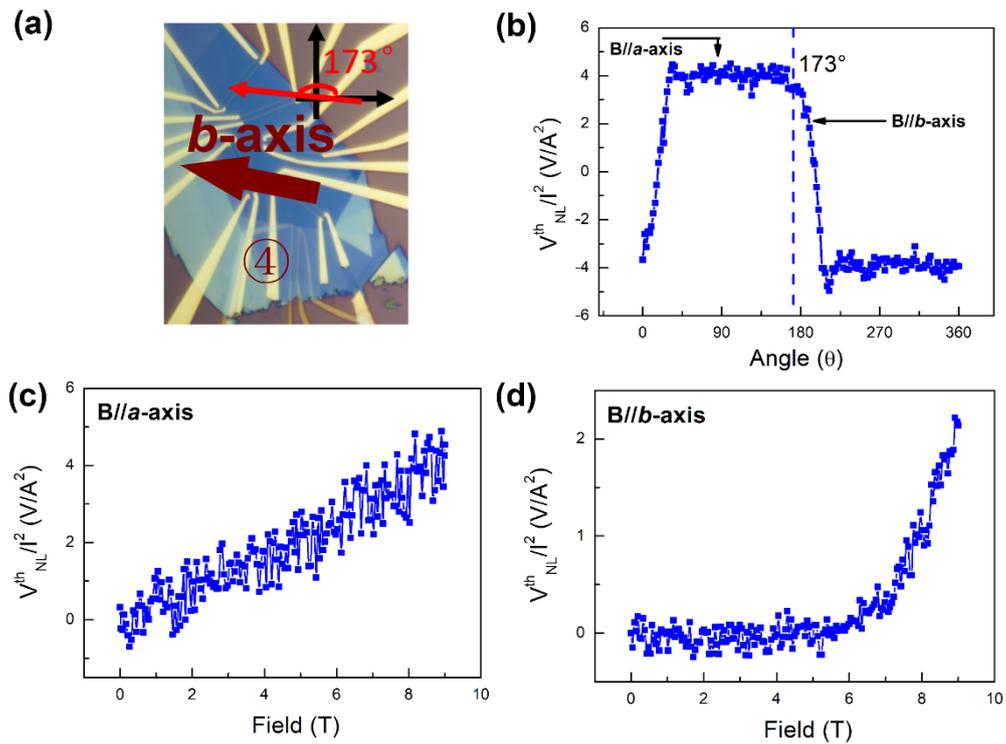

**Figure S5**. (a) Optical image of the nonlocal device with Pt strips aligned along the *a*-axis of NiPS$_3$ flake. (b) Angular dependence of the nonlocal thermally excited magnon signal measured at 20 K. (c)-(d) Field-dependent nonlocal magnon signals for the magnetic field applied along the *a*-axis (94°) (c) and *b*-axis (185°) (d) of the NiPS$_3$ flake, where an upturn corresponding to the spin-flop transition is clearly observed when B//*b*-axis. 0° means that *B* and *I* are perpendicular.

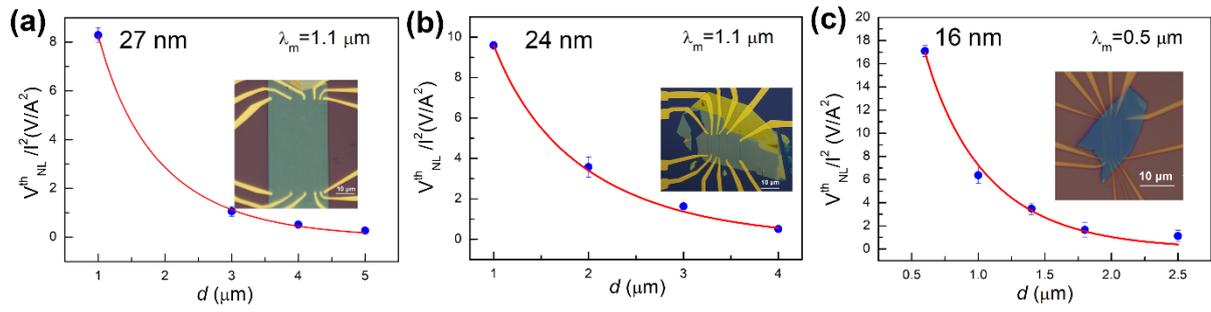

**Figure S6.** (a)-(c) Amplitude of the angle-dependent nonlocal magnon signal as a function of $d$ for different thicknesses of NiPS$_3$ flakes, where the magnon diffusion length ($\lambda_m$) values are extracted by the typical exponential function described in reference (9).

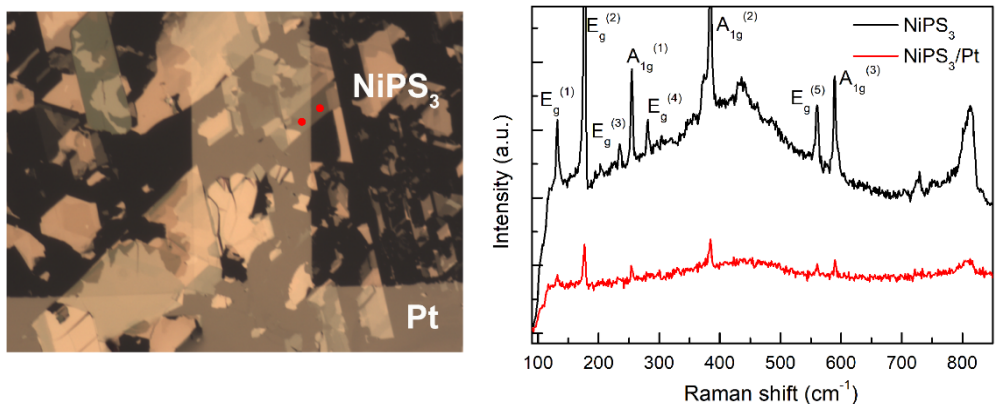

**Figure S7.** Optical image of the NiPS3 flake with a top Pt film (left figure) and the Raman spectra of pristine NiPS3 region and NiPS3/Pt region (right figure). The red spots indicate the measurement positions.